\documentclass[11pt,journal,twoside,onecolumn,draftclsnofoot]{IEEEtran}
\usepackage{cite}
\usepackage{amsmath,amssymb,amsthm,multirow,dsfont,cuted}
\usepackage{enumitem}
\usepackage{epsfig,graphicx}
\usepackage{tikz,pgfplots}
\usepackage{algorithm,algorithmic} 
\usepackage{diagbox}
\usepackage{extarrows}
\usepackage{balance} 
\usepackage{subfigure}
\usepackage{soul}

\newtheorem{theorem}{Theorem}

\newtheorem{assumption}{Assumption}
\newtheorem{lemma}{Lemma}

\newcommand{\cO}{\mathcal{O}}
\newcommand{\N}{\mathcal{N}}

\newcommand{\bA}{\boldsymbol{A}}
\newcommand{\bB}{\boldsymbol{B}}

\newcommand{\bI}{\boldsymbol{I}}

\newcommand{\bR}{\boldsymbol{R}}

\newcommand{\br}{\boldsymbol{r}}
\newcommand{\bu}{\boldsymbol{u}}

\newcommand{\bs}{\boldsymbol{s}}
\newcommand{\bw}{\boldsymbol{w}}
\newcommand{\bx}{\boldsymbol{x}}

\newcommand{\bpsi}{\boldsymbol{\psi}}

\newcommand{\bSig}{\boldsymbol{\Sigma}}
\newcommand{\bsig}{\boldsymbol{\sigma}}

\def\One{\mathds{1}}
\def\Zero{\boldsymbol{0}}

\newcommand{\cA}{\boldsymbol{\mathcal{A}}}
\newcommand{\cB}{\boldsymbol{\mathcal{B}}}

\newcommand{\cF}{\boldsymbol{\mathcal{F}}}
\newcommand{\cG}{\boldsymbol{\mathcal{G}}}

\newcommand{\cM}{\boldsymbol{\mathcal{M}}}

\newcommand{\cR}{\boldsymbol{\mathcal{R}}}
\newcommand{\cS}{\boldsymbol{\mathcal{S}}}

\newcommand{\cX}{\boldsymbol{\mathcal{X}}}

\newcommand{\tr}{\text{Tr}}
\newcommand{\vc}{\text{vec}}

\newcommand{\col}{\text{col}}

\newcommand{\bdiag}{\text{bdiag}}

\newcommand{\expec}{\mathbb{E}}

\begin{document}
\title{Diffusion LMS with Communication Delays: Stability and Performance Analysis}

\author{Fei Hua$^{\dag \ddag}$, Roula Nassif$^\S$, \IEEEmembership{Member, IEEE}, C{\'e}dric Richard$^\ddag$, \IEEEmembership{Senior Member, IEEE}, \\
		Haiyan Wang$^\dag$, Ali H. Sayed$^*$, \IEEEmembership{Fellow, IEEE}
		\\ \vspace{0.2cm}
		\small{\linespread{0.2} $^\dag$ School of Marine Science and Technology\\
			Northwestern Polytechnical University, Xi'an  710072, China\\
             fei.hua@oca.eu \hspace{0.5cm}
			hywang@sust.edu.cn}
		\vspace{0.2cm}\\
		\small{\linespread{0.2} $^\ddag$ Laboratoire Lagrange \\
			Universit\'{e} C\^{o}te d'Azur, Observatoire de la C\^{o}te d'Azur, CNRS, Nice 06108, France \\
      cedric.richard@unice.fr}
    \vspace{0.2cm}\\
    \small{\linespread{0.2} $^\S$ Department of Electrical and Computer Engineering \\
    American University of Beirut, Beirut 1107 2020, Lebanon \\
        roula.nassif@aub.edu.lb  }
		\vspace{0.2cm}\\
		\small{\linespread{0.2} $^*$ School of Engineering \\
			Ecole Polytechnique F\'{e}d\'{e}rale de Lausanne, Lausanne  CH-1015, Switzerland \\
			  ali.sayed@epfl.ch}
	}

\maketitle

\begin{abstract}
We study the problem of distributed estimation over adaptive networks where communication delays exist between nodes.  In particular, we investigate the diffusion Least-Mean-Square (LMS) strategy where delayed intermediate estimates (due to the communication channels) are employed during the combination step. One important question is: Do the delays affect the stability condition and  performance? To answer this question, we conduct a detailed performance analysis in the mean and in the mean-square-error sense of the diffusion LMS with delayed estimates.  Stability conditions, transient and steady-state mean-square-deviation (MSD) expressions are provided. One of the main findings is that  diffusion LMS with delays can still converge under the same step-sizes condition of the diffusion LMS without delays. Finally, simulation results illustrate the theoretical findings.
\end{abstract}


\section{Introduction}

\IEEEPARstart{D}{istributed} estimation aims to estimate a parameter vector  of interest through local computation and cooperation among neighboring nodes.  Distributed solutions  are  attractive in terms of robustness, scalability, and communication overheads. There has been extensive work in the literature studying distributed strategies, such as incremental strategies~\cite{nedic2001incremental,blatt2007convergent,rabbat2005quantized}, consensus strategies~\cite{xiao2005scheme,olfati2007consensus,braca2008enforcing}, {and} diffusion strategies~\cite{Cattivelli2010diff,sayed2013diffusion,sayed2014adaptive}.

Most {prior works on} distributed estimation assume that the network is synchronous and {that} there is no noise or delay in communication.   There exist several  {works studying distributed estimation in the presence of asynchronous events or imperfect information exchanges.} For example, the performance of consensus and gossip strategies is investigated in~\cite{tsitsiklis1986distributed,kar2009distributed,aysal2009broadcast,kar2010distributed} {in the presence of} link-failures, switching topology, and noisy links. Single-task  and multi-task diffusion strategies {over  asynchronous networks are also} examined in~\cite{zhao_asynchronous_2015,zhao_asynchronous_2015-1,zhao_asynchronous_2015-2,nassif2016multitask}, while  diffusion LMS over multi-task networks  is investigated in~\cite{nassif2016diffusion} in the presence of noisy links.

The aforementioned works focus on asynchronous events such as link-failures, switching topology, noisy links, and agents turning on and off {randomly}. Another critical challenge in a distributed implementation {is the presence of} communication delays. For instance, in underwater acoustic sensor networks, one of the main characteristics of underwater communication channels is their long propagation delays due to the low speed of sound. It is therefore crucial  to consider distributed solutions that take into account the communication delays.  In~\cite{olfati2004consensus}, the authors  studied {a situation where the agents in the network} seek  average-consensus in the presence of  time delays. It was shown that, in a \emph{deterministic} setting, time delays should be smaller than a threshold to reach consensus. Recently, several other works {have also derived} interesting findings on  consensus strategies  in the presence of communication delays~\cite{Tsianos2011distributed,wang2015cooperative,Wu2018Decentralized,yang2017distributed,hatanaka2018passivity,yi2019asynchronous}. {In~\cite{Wu2018Decentralized}, the authors propose and analyze the decentralized  asynchronous primal-dual algorithm using fixed step-sizes which  converges to the exact solution. In~\cite{yang2017distributed}, a distributed consensus algorithm is proposed and analyzed for continuous-time multiagent systems in the presence of  communication delays.  Proportional-Integral (PI)  consensus-based distributed optimization algorithm is presented to handle  constant communication delays in~\cite{hatanaka2018passivity}.  In most cases, these works are problematic for processing  streaming data in an \emph{adaptive} context. It has been shown that diffusion strategies outperform consensus or primal-dual strategies in the \emph{stochastic} setting due to gradient noise~\cite{Tu2012Diffusion,Towfic2015Stability}.  By using  \emph{constant} step-sizes, diffusion strategies  ensure continuous learning and adaptation. In this letter, we carry out a detailed performance analysis of diffusion LMS strategy in the presence of delays and provide stability conditions in the mean and mean-square sense. One of the main findings of this work is that  the communication delays do not affect the convergence condition of adaptive diffusion networks.}

The rest of paper is organized as follow: Data model and diffusion strategies with and without delays are introduced in Section~\ref{sec:2}. Stability condition and stochastic behaviors in the mean and mean-square sense of diffusion LMS with delays are provided in Section~\ref{sec:3}. Simulation results are presented in Section~\ref{sec:4}. Finally, Section~\ref{sec:5} concludes this work.

\medskip
\noindent \emph{Notations:} Normal font letters represent scalars, boldface lowercase letters  represent column vectors, and boldface uppercase letters represent matrices.  The symbols $\One$ and $\bI$ are the vector of all ones and the identify matrix of appropriate size, respectively. We use  symbol  $\Zero$ to denote the vector of all zeros or the  matrix of all zeros with appropriate size.  The~$(\ell,k)$-th entry of a matrix $\bA$ is denoted by  $[\bA]_{\ell k}$.  We use the symbol $\otimes$ to denote the Kronecker product and the symbol $\tr(\cdot)$ to denote the trace operator.  The operator $\vc(\bA)$ forms a column vector obtained by stacking the columns of the matrix $\bA$ on top of one another. The operator $\col\{\cdot\}$ constructs a column  vector by stacking the input entries on top of each other. The symbol $\bdiag\{\cdot\}$ constructs block diagonal matrix from input matrix arguments. The symbol $\|\cdot \|_{\infty}$ denotes the  maximum absolute row sum of a  matrix. The symbol $\|\cdot \|_{b,\infty}$ denotes the block maximum norm of a block matrix. The symbols $\rho(\cdot)$ and $\lambda(\cdot)$ denote the spectral radius and eigenvalues of its matrix argument, respectively.

\section{Data model and diffusion strategies} \label{sec:2}
Consider a network of $N$ agents, labeled with $k=1, \ldots, N$. At each time instant $i \geq 0$, each agent $k$ is assumed to have access to a zero-mean scalar measurement $d_{k,i}\in \mathbb{R}$ and a real-valued regression vector $\bx_{k,i} \in  \mathbb{R}^M $ with positive-definite  covariance matrix $\bR_{k}=\expec \{\bx_{k,i} \bx_{k,i}^\top \}$.  Let $\br_{dx,k} \triangleq \expec\{d_{k,i} \bx_{k,i}\}$. The data $\{d_{k,i},\bx_{k,i}\} $ are assumed to be related via the linear regression model:
\begin{equation}
  d_{k,i}=\bx^\top_{k,i} \bw^* + v_{k,i} \label{eq:model}
\end{equation}
where $\bw^* \in\mathbb{R}^M$ is an unknown parameter vector to be estimated, {and $v_{k,i} \in \mathbb{R}$ is a zero-mean  spatially independent} measurement noise with variance $\sigma^2_{v,k}$. 
{In order} to find an estimate for $\bw^*$, the objective of the network is to {minimize} the cost function $J(\bw):\mathbb{R}^M\rightarrow\mathbb{R}$ given by:
\begin{equation}
\min_{\bw}  {} J(\bw) =\sum_{k=1}^N \expec { (d_{k,i}-\bx^\top_{k,i} {\bw} )^2.}
\end{equation} 
{It can be verified that the solution of the above problem is given by $\bw^*=(\sum_{k=1}^N \bR_k)^{-1} (\sum_{k=1}^N \br_{dx,k})$. This requires the signal statistical information $\bR_k$ and $\br_{dx,k}$, which are rarely available in practice.} To solve the problem in a fully-distributed and adaptive manner, the following adapt-then-combine (ATC) diffusion LMS {strategy can be employed}~\cite{Cattivelli2010diff}:
\begin{subequations}
\label{eq:DLMS}
\begin{align}
    \bpsi_{k,i}&=\bw_{k,i-1}+\mu_k \bx_{k,i}(d_{k,i}-\bx_{k,i}^\top  \bw_{k,i-1})  \label{eq:DLMS-adpt}\\
    \bw_{k,i}&=\sum_{\ell \in \mathcal{N}_k} a_{\ell k} \bpsi_{\ell,i}\label{eq:DLMS-comb} 
\end{align}
\end{subequations}
where $\mathcal{N}_k$ denotes the set of neighbors of agent $k$ including itself {and $\bw_{k,i}$ denotes the estimate of $\bw^*$ at agent $k$ and iteration $i$}. The first step~\eqref{eq:DLMS-adpt} is an adaptation step {where agent $k$ uses} its own data available at time $i$ to update the previous estimate $\bw_{k,i-1}$ to  an intermediate estimate $\bpsi_{k,i}$. {Then, in the combination step \eqref{eq:DLMS-comb}, agent $k$}  convexly combines the intermediate estimates $\bpsi_{\ell,i}$ from its neighbors to obtain $\bw_{k,i}$.   The parameter $\mu_k$ is a positive step-size, {and the combination coefficients}  $\{a_{\ell k}\}$ are non-negative, chosen to satisfy the following conditions:
\begin{equation}
 \sum_{\ell=1}^N a_{\ell k} =1, \text{ and } \,  
 \begin{cases}  
        a_{\ell k} >0, \quad \text{if} \; \ell \in \N_k,\\
        a_{\ell k} =0, \quad \text{otherwise}. 
 \end{cases}\label{eq:combcoeff}
\end{equation}
{The above conditions imply that} the matrix $[\bA]_{\ell k}=a_{\ell k}$ {collecting the parameters} $a_{\ell k}$ is left-stochastic.

Notice that, in the combination step~\eqref{eq:DLMS-comb}, each node $k$ at time $i$ is assumed to have access to the   estimates  $\bpsi_{\ell,i}$ obtained by  its neighbors at the \textit{same} time instant $i$. {This requires} the network to be time synchronized and  each node  should transmit its estimate {to its neighbors}  before the next iteration $i+1$. {In this work, we are interested in scenarios where} there are time delays in information exchange. {In this case, at} time $i$, the estimates $\bpsi_{\ell,i}$ from neighbors are not  available at node $k$.  Instead, previous estimates may just have been  received. In principle, each  node $k$ could pause the adaptation step until receiving  the required timely estimates and delay the processing of data by the network.  Alternatively, we can directly use the delayed information without any additional complexity. {By doing so,} we arrive at    ATC  diffusion LMS with delays:
\begin{subequations}
\label{eq:DLMSdelay}
\begin{align}
  \bpsi_{k,i}&=\bw_{k,i-1}+\mu_k \bx_{k,i}(d_{k,i}-\bx_{k,i}^\top  \bw_{k,i-1}) \label{eq:DLMSdelay:adpt}\\
  \bw_{k,i}&=\sum_{\ell \in \mathcal{N}_k} a_{\ell k} {\bpsi_{\ell,i-\tau_{\ell k}}} \label{eq:DLMSdelay:comb}
\end{align}
\end{subequations}
where the adaptation step~\eqref{eq:DLMSdelay:adpt} is the same as~\eqref{eq:DLMS-adpt}. {However, instead of using the timely estimates $\bpsi_{\ell,i}$,} the combination step~\eqref{eq:DLMSdelay:comb} uses delayed estimates {$\bpsi_{\ell,i-\tau_{\ell k}}$} from neighbors, where $\tau_{\ell k} \geq 0$ is an integer denoting the communication delay from node $\ell$ to $k$. 

\section{Performance analysis} \label{sec:3}
We  now analyze the stability and performance of diffusion LMS with delays~\eqref{eq:DLMSdelay}. Before proceeding, we  introduce the following independence assumption.
\begin{assumption}[Independent regressors] \label{asp:1}
The regression vector $\bx_{k,i}$  arises from a stationary random process that is 
temporally white  and  spatially independent with covariance matrix {$\bR_k=\expec \{\bx_{k,i} \bx_{k,i}^\top \}> 0$}. \hfill $\blacksquare$ 
\end{assumption}
A consequence of Assumption~1 is that we can consider the regressors $\{\bx_{k,i}  \}$ independent of $\bw_{\ell,j}$ for all $\ell$ and $j<i$. Although not true in general, this assumption is commonly employed for analyzing  adaptive filters and networks since it simplifies the derivations without constraining the conclusions. Furthermore, there are extensive results in the adaptive filtering literature indicating that the performance results obtained using this assumption match well the actual performance for sufficiently small step-sizes~\cite{sayed2008adaptive}.
\subsection{Error Recursion}
We introduce the error vectors at  node $k$ {and} time instant $i$: 
\begin{equation}
   \widetilde{\bpsi}_{k,i} \triangleq\bw^*-\bpsi_{k,i}, \;
   \widetilde{\bw}_{k,i}\triangleq\bw^*-\bw_{k,i},
\end{equation}
and collect all error vectors into  network block error vectors:  
\begin{equation}
  \widetilde{\bpsi}_{i} \triangleq\col\{\widetilde{\bpsi}_{1,i}, \dots,\widetilde{\bpsi}_{N,i} \},\; 
  \widetilde{\bw}_{i} \triangleq\col\{\widetilde{\bw}_{1,i}, \dots,\widetilde{\bw}_{N,i}\}.
\end{equation}

Subtracting $\bw^*$ {from} both sides of  the adaptation step~\eqref{eq:DLMSdelay:adpt} and using the data model~\eqref{eq:model}, it can be verified that
\begin{equation}
  \widetilde{\bpsi}_i=(\bI_{MN}-\cM \cR_i)\widetilde{\bw}_{i-1} -\cM \bs_i \label{eq:error-psiw}
\end{equation}
where $\cM$ and $\cR_i$ are  $N\times N$ block diagonal matrices with each block of size $M \times M$, $\bs_i$ is an $N \times 1$ block vector whose entries are of size $M \times 1$ each:
\begin{align}
\cM&\triangleq\bdiag \{ \mu_1 \bI_M, \dots, \mu_N \bI_M \},\\
\cR_i&\triangleq\bdiag \{ \bx_{1,i}\bx_{1,i}^\top, \dots, \bx_{N,i}\bx_{N,i}^\top \}, \\
\bs_i&\triangleq\col \{ {\bx}_{1,i}v_{1,i}, \dots, {\bx}_{N,i}v_{N,i} \}.
\end{align}
It holds that $\cR=\expec \cR_i=\bdiag\{\bR_1,\dots,\bR_N \} $ and $\expec \bs_i= \Zero$.

{Due to the communication delays, the relation between $\widetilde{\bw}_{i}$ and $\widetilde{\bpsi}_{i}$ cannot be obtained by simply  subtracting $\bw^*$ from both sides of the combination step~\eqref{eq:DLMSdelay:comb}. Following the same line of reasoning as in~\cite{lee2012spatio,hua2017penaltyb},} we introduce the following {$N(\Gamma+1) \times 1$} extended network block error vectors {with each block of size $M\times 1$:}
\begin{align}
  \widetilde{\bpsi}_{i}^e \triangleq \col\{&\widetilde{\bpsi}_{1,i}, \dots,\widetilde{\bpsi}_{N,i}, \widetilde{\bpsi}_{1,i-1}, \dots,\widetilde{\bpsi}_{N,i-1},\notag \\
        &\dots,\widetilde{\bpsi}_{1,i-\Gamma}, \dots,\widetilde{\bpsi}_{N,i-\Gamma} \},  \\
  \widetilde{\bw}_{i}^e \triangleq \col\{&\widetilde{\bw}_{1,i}, \dots,\widetilde{\bw}_{N,i}, \widetilde{\bpsi}_{1,i}, \dots,\widetilde{\bpsi}_{N,i},\notag \\
  &\dots,\widetilde{\bpsi}_{1,i-\Gamma+1}, \dots,\widetilde{\bpsi}_{N,i-\Gamma+1} \},
\end{align}
where $\Gamma=\max\{\tau_{\ell k} \}, 1\leq \ell, k\leq N $. For {simplicity}, we let $T \triangleq \Gamma+1$. {Using the fact that the matrix $\bA$ is left-stochastic, and} from~\eqref{eq:DLMSdelay:comb}, we obtain 
\begin{equation}
  \widetilde{\bw}_{i}^e= \cA^e \widetilde{\bpsi}_{i}^e \label{eq:error-wepsie}
\end{equation}
with $\cA^e$ an $MNT \times MNT $ matrix {given by}
\begin{equation}
\cA^e =\begin{bmatrix}
\cA_0^\top & \cA_1^\top & \dots &\cA_\Gamma^\top \\
\multicolumn{3}{c}{\bI_{MN\Gamma}}    & {\Zero_{(MN\Gamma) \times (MN)}}
\end{bmatrix}
\end{equation}
where $\cA_\tau \triangleq \bA_\tau \otimes \bI_M $ ($ \tau=0,1,\dots,\Gamma$) {with} the $(\ell,k)$-th entry {of the $N\times N$ matrix} $\bA_\tau$ is given by
\begin{equation}
[\bA_\tau]_{\ell k}=
\begin{cases}  
  [\bA]_{\ell k}, \, &\text{if} \quad \tau_{\ell k}=\tau,\\
  0, \, &\text{otherwise}. 
\end{cases}
\end{equation}
{Observe} that $\bA=\sum_{\tau=0}^\Gamma \bA_\tau$ and $\cA \triangleq \bA \otimes \bI_M =\sum_{\tau=0}^\Gamma \cA_\tau$. 
Likewise, from~\eqref{eq:error-psiw}, we can derive the extended error vector recursive relation between ${\widetilde{\bpsi}}^e_{i}$ and $\widetilde{\bw}_{i-1}^e$:
\begin{equation}
  \widetilde{\bpsi}_i^e=(\bI_{MNT}- \cR_i^e)\widetilde{\bw}_{i-1}^e -  \bs_i^e   \label{eq:error-psiewe}
\end{equation}
with \begin{equation}
\cR_i^e \triangleq 
 \begin{bmatrix}  \cM \cR_i &\Zero_{(MN) \times (MN\Gamma)}\\ \Zero_{(MN\Gamma) \times (MN)} &\Zero_{(MN\Gamma) \times (MN\Gamma)}  \end{bmatrix}, \, 
\bs_i^e\triangleq\begin{bmatrix}  \cM \bs_i  \\ \Zero_{(MN\Gamma) \times 1}    \end{bmatrix}.
\end{equation} 
By combining \eqref{eq:error-wepsie} and~\eqref{eq:error-psiewe}, {we conclude that the} extended network error $\widetilde{\bw}_i^e$ evolves according to the following  recursion: 
\begin{equation}
  \widetilde{\bw}_i^e= \cB_i \widetilde{\bw}_{i-1}^e-  \cA^e\bs_i^e  \label{eq:recursion}
\end{equation}
where $\cB_i \triangleq  \cA^e (\bI_{MNT}- \cR_i^e)$.
\subsection{Mean-error Behavior}
Taking the expectation {of} both sides of~\eqref{eq:recursion}, using Assumption~\ref{asp:1} and  the fact that $\expec \bs_i=\Zero$, we {arrive at the} mean-error recursion:
\begin{align}
  \expec \widetilde{\bw}_{i}^e&= \cA^e (\bI_{MNT}-\cR^e) \expec \widetilde{\bw}_{i-1}^e = \cB   \expec \widetilde{\bw}_{i-1}^e \label{eq:meanerr} \\
\cR^e&=\expec \cR^e_i=\bdiag \{ \cM\cR, \Zero,\dots, \Zero\}, \\
 \cB&=\cA^e (\bI_{MNT}-\cR^e).   \label{eq:cB}
\end{align}

\begin{lemma} \label{lemma:1}
  Consider the block matrix $\cB$ defined by~\eqref{eq:cB}. The matrix $\cB$ is stable, i.e., its spectral radius {$\rho(\cB)$} is less than 1, when   $\| \bI_{MN}-\cM \cR\|_{b,\infty}<1$ where the notation $\|\cdot \|_{b,\infty}$ denotes the block maximum norm of its argument.\footnote{For more details and properties of the block maximum norm,  refer to~\cite[Appendix~D]{sayed2014diffusion}.}  
\end{lemma}
\begin{IEEEproof}
See Appendix~\ref{apd:A}.
\end{IEEEproof}
\begin{theorem}[Convergence in the mean] \label{th:1}
  Assume {the linear} data model~\eqref{eq:model} and  Assumption~\ref{asp:1} hold. Then, for any initial condition,  algorithm \eqref{eq:DLMSdelay} converges  asymptotically in the mean toward the optimal vector $\bw^*$ if  the step-sizes in $\cM$ are chosen to satisfy:
  \begin{equation}
  0<\mu_k<\frac{2}{\lambda_{\max}( \bR_{ k})}, \qquad k=1, \ldots, N. \label{eq:stability}
  \end{equation}
\end{theorem}
\begin{IEEEproof}
  See Appendix~\ref{apd:B}.
\end{IEEEproof}
Theorem~\ref{th:1} allows us to  conclude that  diffusion  LMS with communication delays will continue to converge in the mean sense under the same step-sizes condition of  the  algorithm  without  delays. {It is worth noting that this work focuses on single-task problems where all  nodes seek to estimate the same parameter vector $\bw^*$ in~\eqref{eq:model}. In this case, the resulting estimates $\bw_{k,i}$ will be unbiased according to Theorem~\ref{th:1}. When heterogeneity in the model exists, diffusion multitask algorithms~\cite{nassif2020multitask,chen2014multitask} and exact diffusion algorithms~\cite{Yuan2019Exact} can be used in order to obtain unbiased estimates.}

\subsection{Variance Relation}
We  now study the mean-square-error behavior.  We consider the  mean-square error vector weighted by a positive semi-definite matrix $\bSig$, i.e., {${\expec}\| \widetilde{\bw}_{i}^e \|_{\bSig}^2 \triangleq {\expec}(\widetilde{\bw}_{i}^e)^\top \bSig \widetilde{\bw}_{i}^e $.}  The freedom in selecting $\bSig$  allow us to derive different performance measures about the network and the nodes. {Evaluating the weighted square measures} on both sides of~\eqref{eq:recursion}, we {get}:
\begin{align}
  \| \widetilde{\bw}_{i}^e \|_{\bSig}^2 =& \| \cB_i \widetilde{\bw}_{i-1}^e-  \cA^e\bs_i^e\|_{\bSig }^2  \notag \\
  =&(\widetilde{\bw}_{i-1}^e)^\top\cB_i^\top \bSig \cB_i \widetilde{\bw}_{i-1}^e + (\cA^e\bs_i^e)^\top \bSig \cA^e\bs_i^e-   (\cB_i \widetilde{\bw}_{i-1}^e)^\top \bSig \cA^e\bs_i^e - (\cA^e\bs_i^e)^\top \bSig \cB_i \widetilde{\bw}_{i-1}^e.\label{eq: equation weighted se}
\end{align}
Taking the expectation of both sides of~\eqref{eq: equation weighted se}, using Assumption~\ref{asp:1} and the fact that the expectations of the last two terms on the right-hand side (RHS) of~\eqref{eq: equation weighted se} are zero, we~{obtain:}
\begin{equation}
 \expec \| \widetilde{\bw}_{i}^e \|_{\bSig}^2 = \expec \| \widetilde{\bw}_{i-1}^e \|_{\bSig^\prime}^2 + \expec \{ (\cA^e\bs_i^e)^\top \bSig \cA^e\bs_i^e \} \label{eq:Evariance}
\end{equation} 
where $\bSig^\prime \triangleq \expec \{ \cB_i^\top \bSig \cB_i \}$.
Let $\bsig\triangleq\vc (\bSig)$ denote the vector obtained by stacking the  columns of  the matrix $\bSig$ on top of each other. {Note that, in the sequel, we will use the notation $\|\cdot \|^2_{\bsig}$  and $\|\cdot\|^2_{\bSig}$ interchangeably  to  refer to the same  quantity.} Considering the following properties of matrices:
\begin{align}
\vc(\bA \bSig \bB)=& (\bB^\top \otimes \bA) \bsig, \label{eq:vecprop} \\
\tr(\bSig \bB)=& [\vc(\bB^\top)]^\top \bsig, \label{eq:Trprop}
\end{align} 
we  find that $\bsig^\prime  \triangleq \vc(\bSig^\prime)= \expec \{\cB_i^\top \otimes \cB_i^\top  \} \bsig.$
 Let $\cF \triangleq \expec \{\cB_i^\top \otimes \cB_i^\top  \} $. 
 {which can} be expressed as:
\begin{align}
\cF= &\expec\{ \cB_i^\top \otimes \cB_i^\top\} \notag \\
= &(\cA^e)^\top \otimes (\cA^e)^\top-(\cA^e)^\top \otimes (\cA^e\cR^e)^\top - \notag \\ &(\cA^e\cR^e)^\top \otimes (\cA^e)^\top + \expec\{(\cA^e\cR_i^e)^\top \otimes (\cA^e\cR_i^e)^\top \}.  \label{eq:cFi}
\end{align}
The evaluation of the last expectation term on the RHS of~\eqref{eq:cFi} requires high-order statistical moments of regression data which are usually unavailable. 
However, we can notice that it depends on the square of step-sizes $\{\mu_k^2\}$. This allows us to continue the analysis by taking this factor into account as was done in other studies~\cite{sayed2008adaptive}. Particularly, it is sufficient for the exposition to focus on the case of sufficiently small step-sizes where terms involving higher powers of the step-sizes $\{\mu_k\}$ can be ignored. In this case, the matrix $\cF$  can be {approximated by}:
\begin{equation}
  \cF \approx \cB^\top \otimes \cB^\top. \label{eq:cFaprx}
\end{equation}
The first item on the RHS of~\eqref{eq:Evariance} can be rewritten as $ \expec \| \widetilde{\bw}_{i-1}^e \|_{\bSig^\prime}^2=\expec \| \widetilde{\bw}_{i-1}^e \|_{\cF \bsig}^2 $. Now we evaluate the second term:
\begin{equation}
  \expec \{ (\cA^e\bs_i^e)^\top \bSig \cA^e\bs_i^e \}=\tr(\bSig\cG)=[\vc(\cG^\top)]^\top \bsig
\end{equation}
where 
\begin{align}
 \cG &\triangleq    \cA^e \cS^e (\cA^e)^\top , \\
 \cS^e&\triangleq \expec \{\bs_i^e  (\bs_i^e)^\top\}=\bdiag\{\cS,\Zero, \dots,\Zero  \}, \label{eq:cse}\\
 \cS &\triangleq \cM\cdot\bdiag\{\sigma_{v,k}^2\bR_k \}_{k=1}^N\cdot \cM. 
\end{align}
Therefore, the variance relation~\eqref{eq:Evariance} can be approximated as 
\begin{equation}
  \expec \| \widetilde{\bw}_{i}^e \|_{\bsig}^2 = \expec \| \widetilde{\bw}_{i-1}^e \|_{\cF \bsig}^2 + [\vc(\cG^\top)]^\top \bsig. \label{eq:var}
\end{equation}	
\begin{theorem}[Mean-square stability]
 Consider the same {settings as} in Theorem~\ref{th:1}.  The diffusion LMS with delays algorithm \eqref{eq:DLMSdelay}  is mean-square stable  if the matrix $\cF$ 
 is stable. {Assuming further that the step-sizes are small enough to justify~\eqref{eq:cFaprx}, this condition is satisfied by sufficiently small positive step-sizes that also satisfy~\eqref{eq:stability}.}
\end{theorem}
\begin{IEEEproof}
  See Appendix~\ref{apd:C}.
\end{IEEEproof}

\subsection{Network Transient and Steady-state MSD}
  Iterating~\eqref{eq:var} starting from $i=0$, we {obtain:}
  \begin{equation}
  \expec \|  \widetilde{\bw}_{i}^e \|_{\bsig}^2=\expec \| \widetilde{\bw}_{-1}^e \|_{\cF^{i+1}\bsig}^2+ [\vc(\cG^\top)]^\top\sum_{t=0}^{i} \cF^t \bsig \label{eq:varrlt}
  \end{equation}
  where $\widetilde{\bw}_{-1}^e = \One_{NT} \otimes {\bw^*} $ is an initial condition by assuming ${\bw_{-1}}=\Zero$.
Comparing  relation~\eqref{eq:varrlt} at time instants $i$ and  $i-1$, we can derive the weighted variance recursion
\begin{equation}
  \expec \|  \widetilde{\bw}_{i}^e \|_{\bsig}^2=\expec \| \widetilde{\bw}_{i-1}^e \|_{ \bsig}^2+   \| \widetilde{\bw}_{-1}^e\|^2_{(\cF-\bI)\cF^i \bsig} + [\vc(\cG^\top)]^\top  \cF^i \bsig.
\end{equation}
Let $\zeta_i \triangleq \frac{1}{N} \expec \| \widetilde{\bw}_{i}  \|^2$ denote the network transient MSD averaged over all nodes at time $i$. By replacing $\bsig$ with $\bar{\bsig}= \vc(\bar{\bSig}), \bar{\bSig}= \bdiag\{\bI_{MN}, \Zero, \dots, \Zero \}$, we find that the network transient MSD evolves according to
\begin{equation} 
  \zeta_i=\zeta_{i-1} + \frac{1}{N} \left(\| \widetilde{\bw}_{-1}^e\|^2_{(\cF-\bI)\cF^i \bar{\bsig}} + [\vc(\cG^\top)]^\top  \cF^i \bar{\bsig}\right) \label{eq:TMSD}
\end{equation} 
with $\zeta_{-1}=\frac{1}{N}(\widetilde{\bw}_{-1}^e)^\top \bar{\bSig}\widetilde{\bw}_{-1}^e $. Notice that the evaluation of \eqref{eq:TMSD} involves {the manipulation of the $(MNT)^2 \times (MNT)^2$} matrix, which will be prohibitive in computing. {However, using property~\eqref{eq:Trprop}}, expression~\eqref{eq:TMSD} can be rewritten as 
\begin{align}
  \zeta_i=\zeta_{i-1} + \frac{1}{N} \tr &\Big(\widetilde{\bw}_{-1}^e (\widetilde{\bw}_{-1}^e)^\top  \big( (\cB^{i+1})^\top \bar{\bSig}\cB^{i+1}-  (\cB^{i})^\top \bar{\bSig}\cB^{i}  \big) +(\cB^{i})^\top \bar{\bSig}\cB^{i}\cG \Big) 
\end{align} 
where the matrix operations ease to the order of $\cO(MNT)$.

The network steady-state MSD is defined as 
$ \zeta^* \triangleq \frac{1}{N}  \lim_{i\rightarrow \infty}\expec \| \widetilde{\bw}_{i}  \|^2$.  Assuming that $\cF$ is stable and that the algorithm converges in the mean-square sense,  taking the limit on both sides of~\eqref{eq:varrlt}, we observe that the first term on the RHS converges to zero. Then, by setting $\bsig=\frac{1}{N} \vc (\bar{\bSig})$ and using  properties~\eqref{eq:vecprop} and~\eqref{eq:Trprop}, we get:
\begin{equation}
  \zeta^*=  \frac{1}{N} \sum_{t=0}^\infty \tr  \left( (\cB^{t})^\top \bar{\bSig}\cB^{t}\cG \right). \label{eq:ssMSDalt}
\end{equation}

\section{Simulations}\label{sec:4}

\begin{figure}[ht!]
  \centering
  \subfigure[Network]{ 
      \centering
      \includegraphics[scale=0.45]{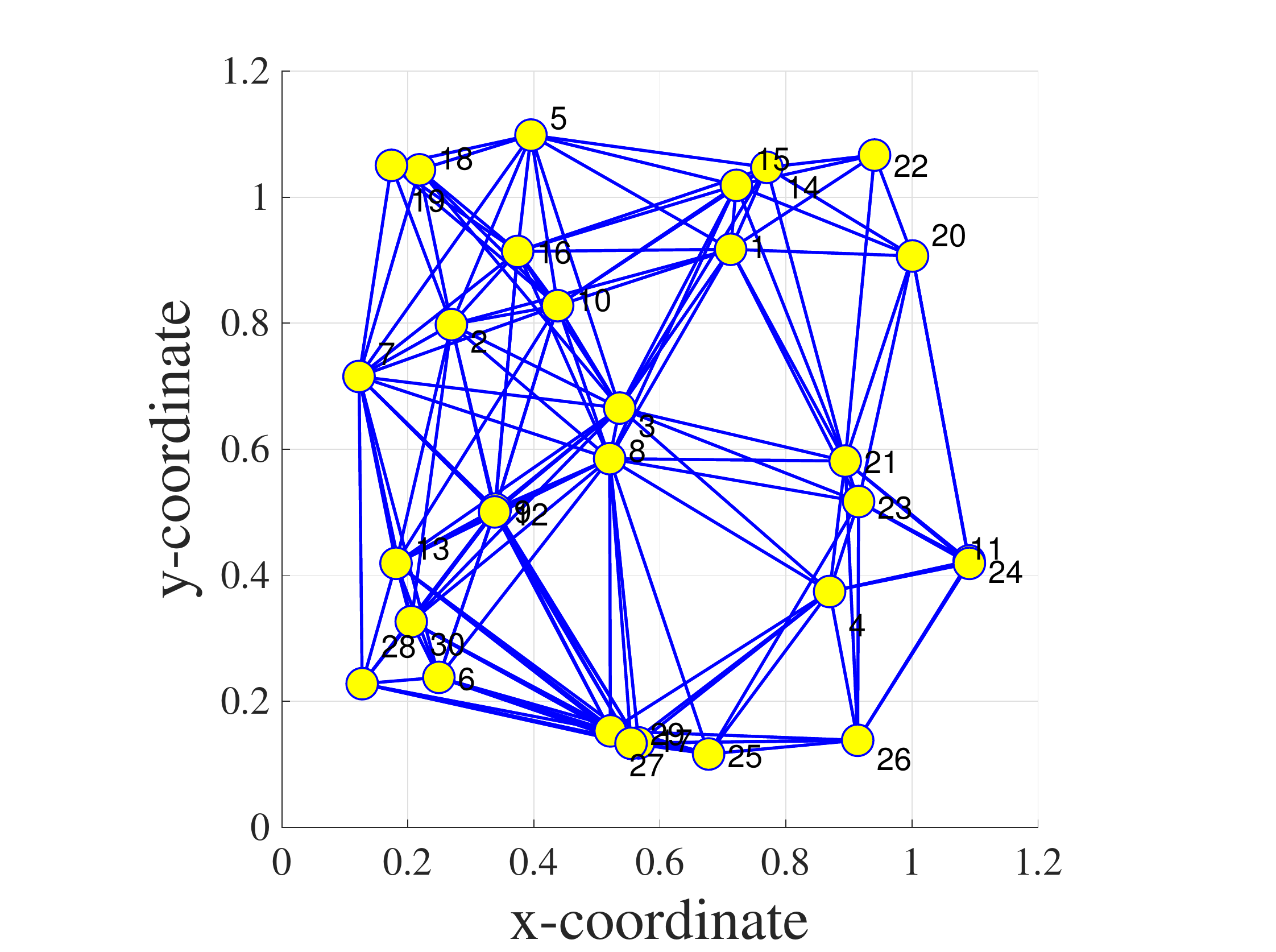}\label{fig:1a} 
  }\\
  \subfigure[Variances]{
      \centering
      \includegraphics[scale=0.4]{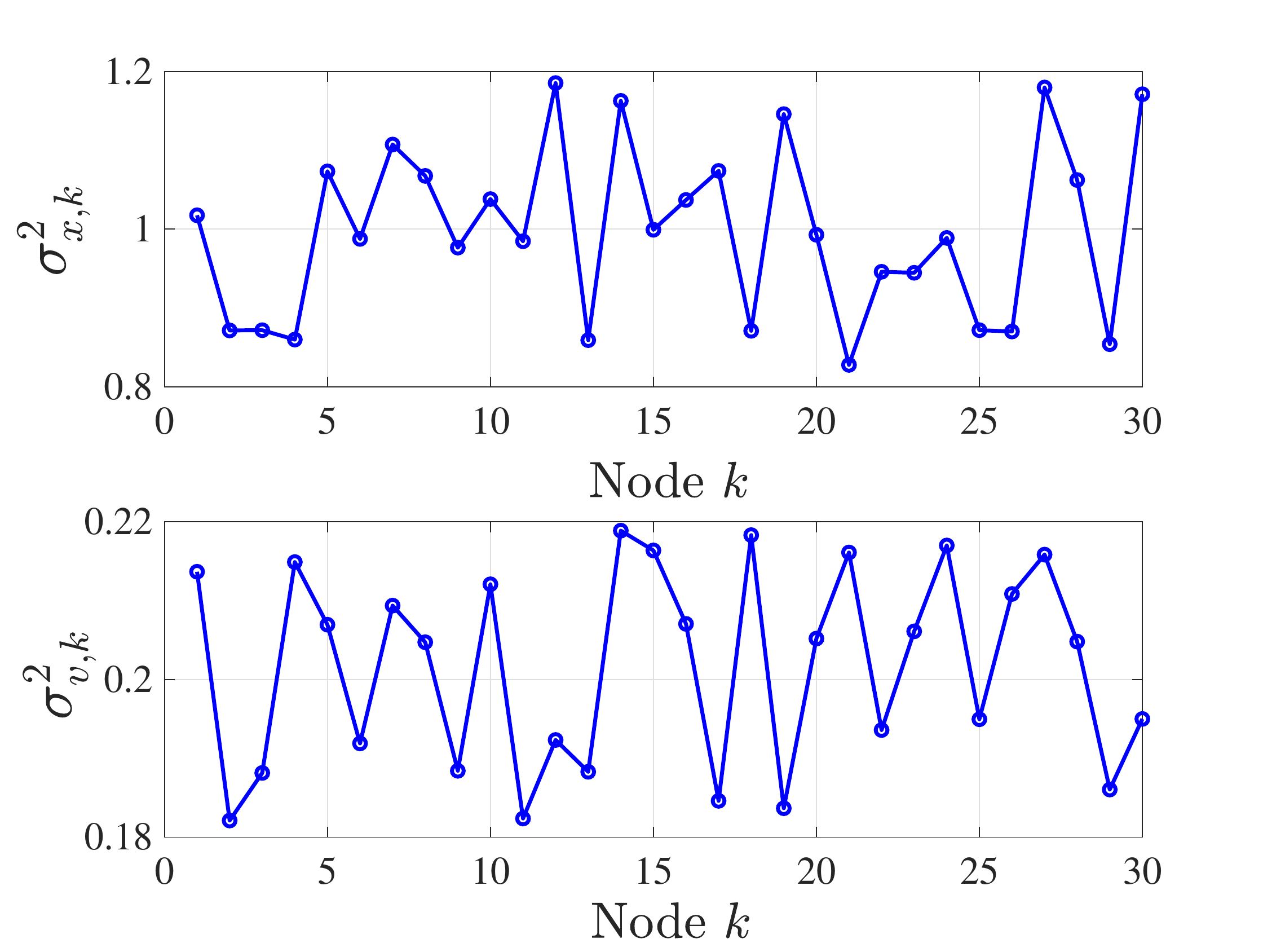}\label{fig:1b} 
  }%
  \caption{Simulation settings. {Variances are generated as $\sigma_{x,k}^2\sim \mathcal{U}(0.8, 1.2)$, $\sigma_{v,k}^2 \sim \mathcal{U}(0.18, 0.22)$.} } \label{fig:1}
\end{figure}

 We consider  a network of 30 nodes with the topology  depicted in Fig.~\ref{fig:1a}. The length of {the parameter} vector is set to $M=10$ and the optimal vector is $\bw^*=[0.495, -0.134, 0.139, -0.328, 0.367, -0.049,\\ -0.141,-1.858,-0.253, -0.602]^\top$. The regression vectors {$\bx_{k,i}$} are generated from a zero-mean Gaussian distribution with covariance matrix $\bR_{k} = \sigma^2_{x,k} \bI_3$. The noises {$v_{k,i}$} are
zero-mean i.i.d. Gaussian random variables with variances $\sigma^2_{v,k}$. The variances $\sigma^2_{x,k}$ and $\sigma^2_{v,k}$ are shown in Fig.~\ref{fig:1b}. The communication delay {$\tau_{\ell k}$} between two connected nodes $\ell$ and $k$ is {proportional} to their distance. The step-sizes are set to  $\mu_k=\mu$ for all nodes. Combination coefficients for diffusion strategies are chosen according to  the uniform rule, i.e., $a_{\ell k}=1/|\mathcal{N}_k|$ for $\ell \in \mathcal{N}_k$. All simulated results are averaged over 500 independent trials.

\begin{figure}[!ht]
 \centering
  \includegraphics[width=0.9\linewidth,height=0.55\linewidth]{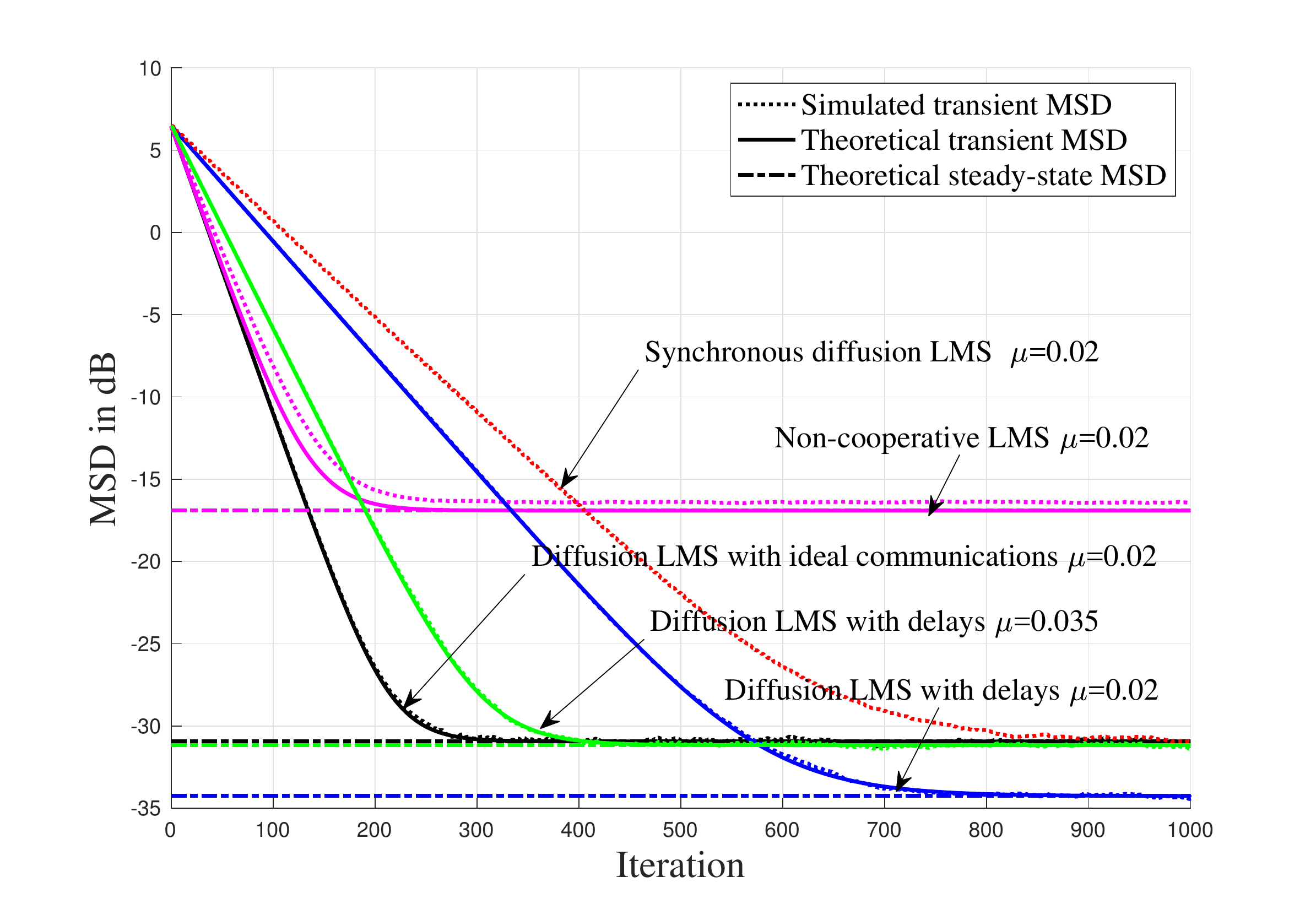}
  \caption{MSD comparison.}\label{fig:2}
\end{figure}

In the first experiment, we set $\mu=0.02$ for non-cooperative LMS, diffusion LMS with  delays, diffusion LMS with ideal communications where there is no delay, and synchronous diffusion LMS. Note that, for the synchronous diffusion LMS,  all nodes need to wait for the longest delayed information to complete one adaptation and combination process.   Observe from Fig.~\ref{fig:2} that the diffusion LMS strategies perform better than non-cooperative LMS in terms of steady-state MSD, and that  diffusion LMS with delays  achieves  better network steady-state MSD compared to synchronous diffusion LMS at a faster convergence rate  and to diffusion LMS with ideal communications at a slower  rate under the same step-sizes. In the second experiment, for comparison purposes, we set $\mu=0.035$ for diffusion LMS with delays in order to meet the same steady-state MSD. It is seen that diffusion LMS with delays converges much faster than the synchronous counterpart. This implies that one can adjust the step-sizes for  diffusion LMS with delays to obtain faster convergence rate without additional asynchronous, computational and storage overheads. Moreover, the algorithm will be stable as long as the step-sizes are small enough,  which  is independent of delays.   Finally, from Fig.~\ref{fig:2}, {we observe that}  the simulated results match well the theoretical~curves.
\section{Conclusion}\label{sec:5}
In this work, we considered the problem of distributed estimation over adaptive networks in the presence of communication delays. We derived  the stability condition for diffusion LMS strategy with delays. Stochastic behaviors in the mean and mean-square sense  were also provided. Simulation results confirmed the theoretical findings.
\appendices
\section{Proof of Lemma 1} \label{apd:A}
Consider the matrix $\cB=\cA^e (\bI_{MNT}-\cR^e)$ defined by~\eqref{eq:cB} and {assume that $\| \bI_{MN}-\cM \cR\|_{b,\infty}<1$}. {Since each row of $\cA^e$ adds up to one and its entries are non-negative, the matrix}  $\cA^e$ is a right-stochastic matrix. It then holds from Lemma~D.4 in~\cite{sayed2014diffusion} that $\| \cA^e\|_{b,\infty}=1$. Since, by assumption,  $\| \bI_{MN}-\cM \cR\|_{b,\infty}<1$, it also holds that the block diagonal matrix $\bI_{MNT}-\cR^e$ {(whose first block is $\bI_{MN}-\cM \cR$ and whose remaining blocks are identity matrices) satisfies} $\|\bI_{MNT}-\cR^e \|_{b,\infty}=1$. {Since} the  spectral radius of a matrix is upper bounded by   any of its induced norms, {and using the sub-multiplicative property of the block maximum norm}, we obtain:
\begin{align}
  \rho(\cB) &\leq \| \cA^e(\bI_{MNT}-\cR^e) \|_{b,\infty} \\ & 
  \leq \| \cA^e \|_{b,\infty} \cdot \|\bI_{MNT}-\cR^e \|_{b,\infty}=1.
\end{align}
Next, we verify that actually $\rho(\cB)$ is strictly less than 1. Since $\lambda(\bA \bB)=\lambda(\bB \bA)$, we have $\rho(\cB)=\rho(\cB^\prime)$ with $\cB^\prime$ given by:
  \begin{align}
  \cB^\prime&=(\bI_{MNT}-\cR^e) \cA^e  =\begin{bmatrix} 
    \cX_0&\dots  &\cX_\Gamma  \\
       \multicolumn{2}{c}{ \bI_{MN\Gamma}}    & \Zero
  \end{bmatrix}, \\
\cX_\tau&=(\bI_{MN}-\cM\cR)\cA_\tau^\top, \quad \tau=0,\dots,\Gamma.
\end{align}
It holds that $\sum_{\tau=0}^\Gamma \cX_\tau=(\bI_{MN}-\cM \cR) \cA^\top$. {Now, let us assume} that $\cB^\prime$ has an eigenvalue $\lambda$ whose magnitude is 1, i.e.,~$|\lambda|=1$, {with} the corresponding eigenvector $\bu=\col\{\bu_0,\dots,\bu_\Gamma\}$ {where} each entry $\bu_\tau$ is of size {$MN\times 1$}. Then, we {can write:}
 \begin{equation}
   \cB^\prime \bu= e^{j\theta} \bu.
 \end{equation}
By expanding the above equation, we obtain:
 \begin{equation}
 \col\left\{\sum_{\tau=0}^\Gamma \cX_\tau \bu_\tau, \bu_0,\dots,\bu_{\Gamma-1} \right\}=e^{j\theta}\col\{ \bu_0,\dots,\bu_{\Gamma}\},
 \end{equation}
 {which gives rise to the following set of equations:}
 \begin{equation}
 \label{eq: set of equation}
 \begin{cases}
 \bu_{\Gamma-1} = e^{j\theta} \bu_\Gamma \\
 \bu_{\Gamma-2} = e^{j\theta} \bu_{\Gamma-1}=e^{j 2 \theta}\bu_{\Gamma}\\
 \vdots\\
 \bu_{0} = e^{j\theta} \bu_1= e^{j \Gamma \theta}\bu_{\Gamma} \\
 \sum_{\tau=0}^\Gamma \cX_\tau  \bu_\tau = e^{j\theta} \bu_0= e^{j (\Gamma+1) \theta}\bu_{\Gamma}\\
 \end{cases}
 \end{equation}
Using the previous equations relating  $\bu_{\tau}$ to $\bu_\Gamma$, substituting them into the last equation in~\eqref{eq: set of equation}, and multiplying both sides of the resulting equation by $e^{-j(\Gamma+1)\theta}$, we obtain: 
 \begin{equation}
   \left( \sum_{\tau=0}^\Gamma \cX_\tau  e^{-j(\tau+1)\theta} \right)\bu_\Gamma=\bu_\Gamma. \label{eq:eigvceq}
 \end{equation}
{Recall that $\| \bI_{MN}-\cM \cR\|_{b,\infty}<1$ and equation~\eqref{eq:eigvceq} followed from assuming that the matrix $\cB^\prime$ has an eigenvalue $\lambda$ whose magnitude is $1$. In the following, we  show that when $\| \bI_{MN}-\cM \cR\|_{b,\infty}<1$, relation~\eqref{eq:eigvceq} cannot be true thus $\cB^\prime$ cannot have an eigenvalue  with magnitude  $1$. In particular, we  show that when $\| \bI_{MN}-\cM \cR\|_{b,\infty}<1$ the spectral radius of the matrix $ \left( \sum_{\tau=0}^\Gamma \cX_\tau  e^{-j(\tau+1)\theta} \right)$ would be strictly less than one, which contradicts~\eqref{eq:eigvceq}. To see this, note that:}
 \begin{align}
   \Big\|\sum_{\tau=0}^\Gamma \cX_\tau &  e^{-j(\tau+1)\theta}\Big\|_{b,\infty} \notag\\ & 
   = \Big\|(\bI_{MN}-\cM \cR) \cdot \sum_{\tau=0}^\Gamma \cA_\tau^\top  e^{-j(\tau+1)\theta}\Big\|_{b,\infty} \notag\\ &
   \overset{(a)}{<} \Big\| \sum_{\tau=0}^\Gamma \cA_\tau^\top  e^{-j(\tau+1)\theta}\Big\|_{b,\infty} \notag \\ &
   \triangleq    \| \underline{\cA}^\top  \|_{b,\infty}  \notag\\ & 
   \overset{(b)}{=}   \| \underline{\bA}^\top \|_{\infty}  \notag\\ & 
   \overset{(c)}{=}   \| \bA^\top \|_{\infty} =1   
 \end{align} 
where $(a)$ follows from the sub-multiplicative property of norms and the fact that $\| \bI_{MN}-\cM \cR\|_{b,\infty}<1$, the entries of $\underline{\cA}$ equal to the entries of $\cA$ at the same positions  scaled by $e^{-j(\tau+1)\theta}$ (this also applies to $\underline{\bA}$ and $\bA$), $(b)$ follows from Lemma D.3 in~\cite{sayed2014diffusion} with  $\| \cdot\|_\infty$ denoting the maximum absolute row sum of its matrix argument, and $(c)$ follows from the fact that  $e^{-j(\tau+1)\theta}$ does not affect the absolute quantity and that $\bA^\top$ is right-stochastic. We therefore conclude that $\rho(\sum_{\tau=0}^\Gamma \cX_\tau  e^{-j(\tau+1)\theta}) <1 $, which contradicts~\eqref{eq:eigvceq}. {This allows us to conclude that the matrix  $\cB^\prime$ cannot have an eigenvalue  with magnitude  $1$, and therefore,} the spectral radius of $\cB$ is less than one.

\section{Proof of Theorem 1}\label{apd:B}
Provided that $\cB$ is stable, the mean-error in~\eqref{eq:meanerr} will converge  to zero. According to Lemma~1, a sufficient condition for ensuring the stability of $\cB$ is to select $\mu_k$ such that $\| \bI_{MN}-\cM \cR\|_{b,\infty}<1$. Following~\cite{sayed2014diffusion}, it can be verified that condition~\eqref{eq:stability} ensures $\| \bI_{MN}-\cM \cR\|_{b,\infty}<1$.
\section{Proof of Theorem 2}\label{apd:C}
  Iterating~\eqref{eq:var} starting from $i=0$, we {obtain:}
  \begin{equation}
  \expec \|  \widetilde{\bw}_{i}^e \|_{\bsig}^2=\expec \| \widetilde{\bw}_{-1}^e \|_{\cF^{i+1}\bsig}^2+ [\vc(\cG^\top)]^\top\sum_{t=0}^{i} \cF^t \bsig 
  \end{equation}
  where $\widetilde{\bw}_{-1}^e = \One_{NT} \otimes {\bw^*} $ is an initial condition by assuming ${\bw_{-1}}=\Zero$. It is easy to verify that   $\cF^{i+1}\bsig $ converges to zero  and the series $\sum_{t=0}^i \cF^t \bsig$ is a bounded vector  as $i \rightarrow \infty$ if $\cF$ is stable. To this end, the diffusion LMS with delays~\eqref{eq:DLMSdelay} is stable in the mean-square sense and its  weighted mean-square error $\expec \|  \widetilde{\bw}_{i}^e \|_{\bsig}^2$ converges to a finite value. Under the  sufficiently small  step-sizes {assumption}, $\cF$ can be evaluated by~\eqref{eq:cFaprx}. In this case, we get $\rho(\cF)=[\rho(\cB)]^2$, and therefore  $\cF$ will be stable if $\cB$ is stable. According to the proof of Theorem~1, $\cB$ is stable if condition~\eqref{eq:stability} is satisfied. Thus, condition~\eqref{eq:stability} ensures mean-square stability of the  algorithm for  sufficiently small  step-sizes. 
\bibliographystyle{IEEEtran}
\bibliography{IEEEabrv,ref}

\begin{thebibliography}{10}
\providecommand{\url}[1]{#1}
\csname url@samestyle\endcsname
\providecommand{\newblock}{\relax}
\providecommand{\bibinfo}[2]{#2}
\providecommand{\BIBentrySTDinterwordspacing}{\spaceskip=0pt\relax}
\providecommand{\BIBentryALTinterwordstretchfactor}{4}
\providecommand{\BIBentryALTinterwordspacing}{\spaceskip=\fontdimen2\font plus
\BIBentryALTinterwordstretchfactor\fontdimen3\font minus
  \fontdimen4\font\relax}
\providecommand{\BIBforeignlanguage}[2]{{%
\expandafter\ifx\csname l@#1\endcsname\relax
\typeout{** WARNING: IEEEtran.bst: No hyphenation pattern has been}%
\typeout{** loaded for the language `#1'. Using the pattern for}%
\typeout{** the default language instead.}%
\else
\language=\csname l@#1\endcsname
\fi
#2}}
\providecommand{\BIBdecl}{\relax}
\BIBdecl

\bibitem{nedic2001incremental}
A.~Nedic and D.~P. Bertsekas, ``Incremental subgradient methods for
  nondifferentiable optimization,'' \emph{SIAM J. Optim.}, vol.~12, no.~1, pp.
  109--138, 2001.

\bibitem{blatt2007convergent}
D.~Blatt, A.~O. Hero, and H.~Gauchman, ``A convergent incremental gradient
  method with a constant step size,'' \emph{SIAM J. Optim.}, vol.~18, no.~1,
  pp. 29--51, 2007.

\bibitem{rabbat2005quantized}
M.~G. Rabbat and R.~D. Nowak, ``Quantized incremental algorithms for
  distributed optimization,'' \emph{{IEEE} J. Sel. Areas Commun.}, vol.~23,
  no.~4, pp. 798--808, Apr. 2005.

\bibitem{xiao2005scheme}
L.~{Xiao}, S.~{Boyd}, and S.~{Lall}, ``A scheme for robust distributed sensor
  fusion based on average consensus,'' in \emph{Proc. 2005 4th Int. Symp. Inf.
  Process. Sensor Netw.}, Boise, ID, USA, 2005, pp. 63--70.

\bibitem{olfati2007consensus}
R.~Olfati-Saber, J.~A. Fax, and R.~M. Murray, ``Consensus and cooperation in
  networked multi-agent systems,'' \emph{Proc. {IEEE}}, vol.~95, no.~1, pp.
  215--233, Jan. 2007.

\bibitem{braca2008enforcing}
P.~Braca, S.~Marano, and V.~Matta, ``Enforcing consensus while monitoring the
  environment in wireless sensor networks,'' \emph{{IEEE} Trans. Signal
  Process.}, vol.~56, no.~7, pp. 3375--3380, Jun. 2008.

\bibitem{Cattivelli2010diff}
F.~S. Cattivelli and A.~H. Sayed, ``Diffusion {LMS} strategies for distributed
  estimation,'' \emph{{IEEE} Trans. Signal Process.}, vol.~58, no.~3, pp.
  1035--1048, Mar. 2010.

\bibitem{sayed2013diffusion}
A.~H. Sayed, S.-Y. Tu, J.~Chen, X.~Zhao, and Z.~J. Towfic, ``Diffusion
  strategies for adaptation and learning over networks,'' \emph{{IEEE} Signal
  Process. Mag.}, vol.~30, no.~3, pp. 155--171, May. 2013.

\bibitem{sayed2014adaptive}
A.~H. Sayed, ``Adaptive networks,'' \emph{Proc. {IEEE}}, vol. 102, no.~4, pp.
  460--497, Apr. 2014.

\bibitem{tsitsiklis1986distributed}
J.~Tsitsiklis, D.~Bertsekas, and M.~Athans, ``Distributed asynchronous
  deterministic and stochastic gradient optimization algorithms,'' \emph{{IEEE}
  Trans. Autom. Control}, vol.~31, no.~9, pp. 803--812, Sep. 1986.

\bibitem{kar2009distributed}
S.~Kar and J.~M.~F. Moura, ``Distributed consensus algorithms in sensor
  networks with imperfect communication: Link failures and channel noise,''
  \emph{{IEEE} Trans. Signal Process.}, vol.~57, no.~1, pp. 355--369, Jan.
  2009.

\bibitem{aysal2009broadcast}
T.~C. Aysal, M.~E. Yildiz, A.~D. Sarwate, and A.~Scaglione, ``Broadcast gossip
  algorithms for consensus,'' \emph{{IEEE} Trans. Signal Process.}, vol.~57,
  no.~7, pp. 2748--2761, Jul. 2009.

\bibitem{kar2010distributed}
S.~Kar and J.~M.~F. Moura, ``Distributed consensus algorithms in sensor
  networks: Quantized data and random link failures,'' \emph{{IEEE} Trans.
  Signal Process.}, vol.~58, no.~3, pp. 1383--1400, Mar. 2010.

\bibitem{zhao_asynchronous_2015}
X.~Zhao and A.~H. Sayed, ``Asynchronous adaptation and learning over
  networks--{Part I}: Modeling and stability analysis,'' \emph{{IEEE} Trans.
  Signal Process.}, vol.~63, no.~4, pp. 811--826, Feb. 2015.

\bibitem{zhao_asynchronous_2015-1}
------, ``Asynchronous adaptation and learning over networks--{Part II}:
  Performance analysis,'' \emph{{IEEE} Trans. Signal Process.}, vol.~63, no.~4,
  pp. 827--842, Feb. 2015.

\bibitem{zhao_asynchronous_2015-2}
------, ``Asynchronous adaptation and learning over networks--{Part III}:
  Comparison analysis,'' \emph{{IEEE} Trans. Signal Process.}, vol.~63, no.~4,
  pp. 843--858, Feb. 2015.

\bibitem{nassif2016multitask}
R.~Nassif, C.~Richard, A.~Ferrari, and A.~H. Sayed, ``Multitask diffusion
  adaptation over asynchronous networks,'' \emph{{IEEE} Trans. Signal
  Process.}, vol.~64, no.~11, pp. 2835--2850, Jun. 2016.

\bibitem{nassif2016diffusion}
R.~Nassif, C.~Richard, J.~Chen, A.~Ferrari, and A.~H. Sayed, ``Diffusion {LMS}
  over multitask networks with noisy links,'' in \emph{Proc. IEEE Int. Conf.
  Acoust., Speech, Signal Process.}, Shanghai, China, May 2016, pp. 4583--4587.

\bibitem{olfati2004consensus}
R.~Olfati-Saber and R.~Murray, ``Consensus problems in networks of agents with
  switching topology and time-delays,'' \emph{{IEEE} Trans. Autom. Control},
  vol.~49, no.~9, pp. 1520--1533, Sep. 2004.

\bibitem{Tsianos2011distributed}
K.~I. {Tsianos} and M.~G. {Rabbat}, ``Distributed consensus and optimization
  under communication delays,'' in \emph{Proc. 49th Annu. Allerton Conf.
  Commun. Control Comput.}, Monticello, IL, USA, Sep. 2011, pp. 974--982.

\bibitem{wang2015cooperative}
H.~Wang, X.~Liao, T.~Huang, and C.~Li, ``Cooperative distributed optimization
  in multiagent networks with delays,'' \emph{{IEEE} Trans. Syst., Man,
  Cybern., Syst.}, vol.~45, no.~2, pp. 363--369, Feb. 2015.

\bibitem{Wu2018Decentralized}
T.~{Wu}, K.~{Yuan}, Q.~{Ling}, W.~{Yin}, and A.~H. {Sayed}, ``Decentralized
  consensus optimization with asynchrony and delays,'' \emph{{IEEE} Trans.
  Signal Inf. Process. Netw.}, vol.~4, no.~2, pp. 293--307, Jun. 2018.

\bibitem{yang2017distributed}
S.~Yang, Q.~Liu, and J.~Wang, ``Distributed optimization based on a multiagent
  system in the presence of communication delays,'' \emph{{IEEE} Trans. Syst.,
  Man, Cybern., Syst.}, vol.~47, no.~5, pp. 717--728, Mar. 2017.

\bibitem{hatanaka2018passivity}
T.~Hatanaka, N.~Chopra, T.~Ishizaki, and N.~Li, ``Passivity-based distributed
  optimization with communication delays using {PI} consensus algorithm,''
  \emph{{IEEE} Trans. Autom. Control}, vol.~63, no.~12, pp. 4421--4428, Apr.
  2018.

\bibitem{yi2019asynchronous}
P.~{Yi} and L.~{Pavel}, ``Asynchronous distributed algorithms for seeking
  generalized nash equilibria under full and partial-decision information,''
  \emph{{IEEE} Trans. Cybern.}, pp. 1--13, to be published, 2019.

\bibitem{Tu2012Diffusion}
S.-Y. Tu and A.~H. Sayed, ``Diffusion strategies outperform consensus
  strategies for distributed estimation over adaptive networks,'' \emph{{IEEE}
  Trans. Signal Process.}, vol.~60, no.~12, pp. 6217--6234, Dec. 2012.

\bibitem{Towfic2015Stability}
Z.~J. Towfic and A.~H. Sayed, ``Stability and performance limits of adaptive
  primal-dual networks,'' \emph{{IEEE} Trans. Signal Process.}, vol.~63,
  no.~11, pp. 2888--2903, Jun. 2015.

\bibitem{sayed2008adaptive}
A.~H. Sayed, \emph{Adaptive Filters}.\hskip 1em plus 0.5em minus 0.4em\relax
  Hoboken, NY, USA: John Wiley \& Sons, 2008.

\bibitem{lee2012spatio}
J.-W. Lee, S.-E. Kim, W.-J. Song, and A.~H. Sayed, ``Spatio-temporal diffusion
  strategies for estimation and detection over networks,'' \emph{{IEEE} Trans.
  Signal Process.}, vol.~60, no.~8, pp. 4017--4034, Aug. 2012.

\bibitem{hua2017penaltyb}
F.~Hua, R.~Nassif, C.~Richard, and H.~Wang, ``Penalty-based multitask
  estimation with non-local linear equality constraints,'' in \emph{Proc. IEEE
  Int. Workshop Comput. Adv. Multi-Sensor Adapt. Process.}, Curacao,
  Netherlands Antilles, Dec. 2017, pp. 1--5.

\bibitem{sayed2014diffusion}
A.~H. Sayed, ``Diffusion adaptation over networks,'' in \emph{Academic Press
  Library in Signal Processing}, S.~Theodoridis and R.~Chellappa, Eds.\hskip
  1em plus 0.5em minus 0.4em\relax Academic Press, Elsevier, 2014, vol.~3, pp.
  322--454.

\bibitem{nassif2020multitask}
R.~Nassif, S.~Vlaski, C.~Richard, J.~Chen, and A.~H. Sayed, ``Multitask
  learning over graphs,'' \emph{\emph{to appear in} IEEE Signal Process. Mag.
  \emph{Also available as arXiv:2001.02112}}, May 2020.

\bibitem{chen2014multitask}
J.~Chen, C.~Richard, and A.~H. Sayed, ``Multitask diffusion adaptation over
  networks,'' \emph{{IEEE} Trans. Signal Process.}, vol.~62, no.~16, pp.
  4129--4144, Aug. 2014.

\bibitem{Yuan2019Exact}
K.~Yuan, B.~Ying, X.~Zhao, and A.~H. Sayed, ``Exact diffusion for distributed
  optimization and learning--{Part I}: Algorithm development,'' \emph{{IEEE}
  Trans. Signal Process.}, vol.~67, no.~3, pp. 708--723, Feb. 2019.

\end{thebibliography}

\end{document}